\documentclass[10pt]{article}
 
\usepackage{fullpage}
\usepackage[all=normal,bibliography=tight]{savetrees}
\usepackage{microtype}
\usepackage{todonotes}
\usepackage{amsthm}
\usepackage{amssymb,bbm}
\usepackage{microtype}
\usepackage{xspace}
\usepackage{todonotes}
\usepackage{comment}
\usepackage{enumerate}
\usepackage{tikz}
\usepackage{graphicx}
\usepackage[absolute]{textpos}
\usetikzlibrary{arrows, decorations.markings}
\usetikzlibrary{fit}					
\usetikzlibrary{backgrounds}

\usepackage{amsmath,amssymb,amsthm}
\usepackage{graphicx}
\usepackage{caption}
\usepackage{subcaption}

\newcommand{\uselipics}{no}

\usepackage{multirow}

\newcommand{\iflipics}[2]{\ifthenelse{\equal{\uselipics}{yes}}{#1}{#2}}

\def\cqedsymbol{\ifmmode$\lrcorner$\else{\unskip\nobreak\hfil
\penalty50\hskip1em\null\nobreak\hfil$\lrcorner$
\parfillskip=0pt\finalhyphendemerits=0\endgraf}\fi}

\newcommand{\Oh}{\ensuremath{\mathcal{O}}}

\usepackage{url}

\numberwithin{equation}{section}

\theoremstyle{definition}

\title{Experimental Evaluation of Parameterized Algorithms for Feedback Vertex Set%
\thanks{Supported by the ``Recent trends in kernelization: theory and experimental evaluation'' project, carried out within the Homing programme of the Foundation for Polish Science co-financed by the European Union under the European Regional Development Fund.}}

\author{Krzysztof Kiljan\thanks{Institute of Informatics, University of Warsaw, Poland. \texttt{krzysztof.kiljan@student.uw.edu.pl}}
  \and
    Marcin Pilipczuk\thanks{Institute of Informatics, University of Warsaw, Poland. \texttt{malcin@mimuw.edu.pl}}}

\date{}

\begin{document}

\maketitle

\begin{abstract}
\textsc{Feedback Vertex Set} is a classic combinatorial optimization problem that asks for
a minimum set of vertices in a given graph whose deletion makes the graph acyclic. 
From the point of view of parameterized algorithms and fixed-parameter tractability, \textsc{Feedback Vertex Set} is one of the landmark problems:
a long line of study resulted in multiple algorithmic approaches and deep understanding of the combinatorics of the problem.
Because of its central role in parameterized complexity, the first edition of the Parameterized Algorithms and Computational Experiments Challenge (PACE) in 2016
featured \textsc{Feedback Vertex Set} as the problem of choice in one of its tracks. 
The results of PACE 2016 on one hand showed large discrepancy between performance of different classic approaches to the problem,
and on the other hand indicated a new approach based on half-integral relaxations of the problem as probably the most efficient approach to the problem.
In this paper we provide an exhaustive experimental evaluation of fixed-parameter and branching algorithms for \textsc{Feedback Vertex Set}.
\end{abstract}

\section{Introduction}

The \textsc{Feedback Vertex Set} problem asks to delete from a given graph a minimum number of vertices to make it acyclic.
It is one of the classic graph optimization problems, appearing already on Karp's list of 21 NP-hard problems~\cite{karp}.
In this work, we are mostly focusing on \emph{fixed-parameter algorithms} for the problem, that is, exact (and thus exponential-time,
as we are dealing with an NP-hard problem) algorithms whose exponential blow-up in the running time bound is confined
by a proper parameterization. More formally, a fixed-parameter algorithm on an instance of size $n$ with parameter value $k$
runs in time bounded by $f(k) \cdot n^c$ for some computable (usually exponential) function $f$ and a constant $c$ independent of $k$.
 
\textsc{Feedback Vertex Set} is one of the most-studied problems from the point of view of parameterized algorithms.
A long line of research~\cite{fvs1,fvs2,fvs3,fvs5,fvs6,fvs7,guo:fvs,fvs:5k,fvs:3.83k,fvs:4krand,fvs:3k}
lead to a very good understanding of the combinatorics of the problem, multiple known algorithmic approaches,
and a long ``race'' for the fastest parameterized algorithm under the parameterization of the solution size (i.e., the parameter $k$
equals the size of the solution we are looking for). Among many approaches, one can find the following:
\begin{enumerate}
\item A classic randomized algorithm of Becker et al.~\cite{fvs:4krand} with expected
running time $4^k n^{\Oh(1)}$.\\
This algorithm is based on the following observation: after performing a set of simple reductions that reduce the graph to minimum degree at least $3$, at least half of the edges of the graph are incident with solution vertices, so a random endpoint of an edge choosen uniformly at random is in the solution with probability at least $1/4$.
\item A line of research of branching algorithms based on iterative compression and an intricate measure 
to bound the size of the branching tree~\cite{fvs:5k,fvs:3.83k,KociumakaP}, leading to a simple $3.62^k n^{\Oh(1)}$-time deterministic algorithm~\cite{KociumakaP}.\\
These algorithms solve a ``disjoint'' version of the problem where, given a set $U \subseteq V(G)$ such that $G-U$ is a forest, one seeks for a solution \emph{disjoint} with $U$ of size at most $k$. A branching algorithm picks a vertex $v \in V(G) \setminus U$ and includes it in the solution (deletes and decreases $k$ by one) or moves to $U$. The crucial observation is that here, apart from $k$, the number of connected components of $G[U]$ is a useful potential to measure the progress of a branching algorithm: if the branching pivot $v$ has many neighbors in $U$, then moving it to $U$ decreases the number of connected components significantly.
\item A polynomial-time algorithm for \textsc{Feedback Vertex Set} in subcubic graphs~\cite{fvs:3.83k} (see also~\cite{KociumakaP} for a simpler proof), 
  used heavily in other approaches as a subroutine.\\
The algorithm build on a reduction from \textsc{Feedback Vertex Set}
in cubic graphs to the matroid matching problem in graphic matroids, which
is polynomial-time solvable.
\item A Monte Carlo algorithm running in time $3^k n^{\Oh(1)}$ via the Cut\&Count technique~\cite{fvs:3k}.\\
The Cut\&Count technique is a generic framework for turning
algorithms with running time $2^{\Oh(t \log t)} n^{\Oh(1)}$ for connectivity
problems in graphs of treewidth bounded by $t$ into Monte-Carlo algorithms
with running time bounds $c^t n^{\Oh(1)}$ by changing the 
``connectivity'' requirement into a modulo-2 counting requirement.
For \textsc{Feedback Vertex Set}, the base of the exponent has been optimized
to $c=3$, and has been proven to be optimal under the
Strong Exponential Time Hypothesis.
\item A surprisingly simple branching algorithm by Cao~\cite{Cao} with a running time bound $8^k n^{\Oh(1)}$.\\
Cao~\cite{Cao} shown that the following algorithm has running time
bounded by $8^k n^{\Oh(1)}$: after applying simple reduction rules,
one either branch on maximum-degree
vertex or, if the maximum degree is at most $3$, solve the instance
in polynomial time.
\item A branching algorithm based on intricate half-integral relaxation due to Imanishi and Iwata~\cite{IwataGithub,Iwata17} with running time bound
$4^k k^{\Oh(1)} n$.\\
The work of Iwata, Wahlstr\"{o}m, and Yoshida~\cite{IwataWY16}
showed a generic framework for branching algorithms for various graph separation
problems. In the case of \textsc{Feedback Vertex Set},~\cite{IwataWY16}
shows that the following local problem has a polynomial-time solvable
half-integral relaxation: given a root vertex $s \in V(G)$, find
a minimum-size set $X \subseteq V(G) \setminus \{s\}$ such that the
connected component of $G-X$ that contains $s$ is a tree. 
Imanishi and Iwata~\cite{IwataGithub,Iwata17} showed a linear-time combinatorial algorithm
that solves this half-integral relaxation. The final algorithm follows the framework of~\cite{IwataWY16}: it branches on vertices for which the half-integral
relaxation solution was undecided (i.e., gave it the non-integral value of $1/2$). 

\item A formulation of a problem as an integer linear program that can be solved using a general ILP solver.
For every vertex $v \in V(G)$ one can create a variable $x_v$ denoting whether we put the given vertex in the solution ($x_v = 1$), or we leave the vertex in the graph ($x_v = 0$). 
This special case of integer linear programming, where every variable is not only an integer, but also binary, is also on Karp's list of 21 NP-hard problems~\cite{karp}.
However, there exist several general integer linear program solvers that often work very efficiently in practice. 
\\ 
Having defined variables for the set of vertices, one can create constraints twofold. 
The first approach is to create a constraint for every cycle, stating that the sum of variables in it has to be at least equal to $1$. It means the same as the definition of the problem --- we need to delete at least one vertex in every cycle. 
\\
Second way to define the problem in terms of integer linear programming is to create constraints for every subset of vertices. For those subsets we want to delete as few vertices as possible to make it a tree. In this approach we need more variables than only for vertices, but the worst-case number of constraints is smaller.
\\
In both of these approaches there is a small drawback -- it is possible that there will be exponential number of constraints. For this reason, the algorithms based on these ideas need to add the constraints dynamically and check whether the solution using only a part of constraints implies the solution for the whole graph.
\end{enumerate}
Another line of research concerns so-called polynomial kernels
for the problem~\cite{fvs:kernel1,fvs:kernel2,fvs:quadratic-kernel,Iwata17}.

Parameterized Algorithms and Computational Experiments challenge is an annual programming challenge started in 2016 that aims 
to ``investigate the applicability of algorithmic ideas studied and developed in the subfields of multivariate, fine-grained, parameterized, or fixed-parameter tractable algorithms''. 
 With three successful editions so far~\cite{PACE2016,PACE2017,PACE2018} and the fourth one just announced, PACE continues to bring together theory and practice in parameterized algorithms community.

The first edition of PACE in 2016 featured \textsc{Feedback Vertex Set}
as the problem of choice in Track B. The winning entry by Imanishi and Iwata,
implementing the aforementioned algorithm based on half-integral relaxation,
turned out to outperform the second entry by the second author of
this work~\cite{PilipczukPACE}, based on the algorithm of Cao~\cite{Cao}. 
The winning margin has been substantial: out of 130 test cases, the winning
entry solved 84, while the second entry solved 66.

These results indicated the branching algorithms based on 
half-integral relaxation of the problem as potentially most efficient 
approach to \textsc{Feedback Vertex Set} in practice.
Furthermore, experimental results of Akiba and Iwata~\cite{AkibaI16}
showed big potential in a branching
algorithm based on the same principle for the \textsc{Vertex Cover} problem.

In the light of the above, we see the need to rigorously experimentally
evaluate different approaches to \textsc{Feedback Vertex Set}.
While the results of PACE 2016 indicate algorithms based on half-integral
relaxation as potentially fastest, a lot of differences may come from 
the use of different preprocessing routines, different choice of lower bounding
or prunning techniques, or even simply different data structures handling basic graph operations.%
\footnote{A good example here is as follows. In \textsc{Feedback Vertex Set}, it is natural to keep the graph in the form of adjacency lists, as the considered graphs are usually of
  constant average degree. However, given an edge $uv$, it is not clear whether the vertex $u$ in its adjacency list should only
    store the vertex $v$, or also a pointer to the position where $v$ keeps $u$ in its adjacency list. On one hand, such pointers greatly simplify the operations of deleting
    a vertex or contracting an edge. On the other hand, they effectively double the size of the graph data structure, increasing the cost of copying the graph in the branching step.}

In this work, we offer such a comparison. We implement a number of
branching strategies mentioned above.
Our implementations use the same data structures for handling graphs,
the same implementations of basic graph operations, 
the same basic reduction rules such as suppressing degree-2 vertices,
and the same branching framework. 
Most of the tested approaches differ only at a small fraction of code: they
usually differ by the choice of the branching pivot, and 
some use one or more approach-specific reduction rules.

In our experiments, we follow up the set up of PACE 2016: we take their
230 instances (100 public and 130 hidden, on which evaluation took place)
as our benchmark set, allow each algorithm to run for 30 minutes on each
test instance.

The paper is organized as follows. In Section~\ref{sec:impl} we discuss the
studied approaches and some technical details of the implementations.
Section~\ref{sec:setup} discusses the setup of the experiment.
Section~\ref{sec:results} presents results, while Section~\ref{sec:conc}
concludes the paper.

\section{Studied algorithms}\label{sec:impl}

Most of the known branching algorithm for \textsc{Feedback Vertex Set}, in particular all algorithms studied in our work, follow the following
general framework.

Every instance to be solved by a recursive branching algorithm constists of a multigraph $G$, a set $U \subseteq V(G)$ of \emph{undeletable} vertices
and allowed budget $k$ for solution size. The goal is to find a set $X \subseteq V(G) \setminus U$ of size at most $k$
such that $G-X$ is a forest.
Each branching step consists of picking a vertex $v \in V(G) \setminus U$ and branching into two cases: either $v$ gets picked to a solution
(and the algorithm recurses on the instance $(G-\{v\}, U, k-1)$) or moved to set $U$ (and the algorithm recurses on the instance $(G, U \cup \{v\}, k)$). 

The intuition of the progress of the algorithm is as follows. In the first branch, the budget $k$ gets decreased. For the second branch,
note that the algorithm can safely terminate for instances with $G[U]$ not being acyclic. Thus, if the branching pivot $v$ has $d$ neighbors in $U$, then
the number of connected components of $G[U \cup \{v\}]$ decreases by $(d-1)$ as compared to $G[U]$.

Between branching step, the algorithm is allowed to perform a number of reduction (preprocessing) steps. 
In the literature, a number of simple reduction steps are known that are performed by all our algorithms:
\begin{enumerate}
\item If $k$ gets negative or $G[U]$ is not acyclic, stop.
\item Remove all vertices of degree at most $1$.
\item If two vertices are connected by more than $2$ parallel edges, reduce their multiplicity to $2$.
\item If there exists a vertex $v$ that has a self-loop, or there exists a single connected component $D$ of $G[U]$ more than one edge incident with $v$ and a vertex of $D$ (i.e., there exists a cycle $C$ in $G$ with $v$ being the only vertex of $V(C) \setminus U$), then greedily include it in the solution
(i.e., delete it and decrease $k$ by one). 
\item Suppress vertices of degree $2$. That is, if a vertex $v$ is of degree $2$ with incident edges $vu$ and $vw$ is present, delete $v$ and replace
it with an edge $uw$. 
\item If there exists a vertex $v$ with two neighbors $u$ and $w$, such that $vu$ is a single edge and $vw$ is a double edge, greedily include $w$ in the solution.
\end{enumerate}
For efficiency, the above reduction rules are implemented in the form of a
queue of vertices to reduce: when the number of distinct neighbors of a vertex
drop to two or less, or a vertex gets a self-loop,
it is enqueued, and preprocessing routines start by clearing up the queue. 

Other preprocessing steps used by some of our algorithms are:
\begin{description}
\item[split into connected components] If the instance becomes disconnected, solve each connected component independently.
Here, we observed that it is often the case that if the current graph is disconnected, then it
contains one large connected component and a few small ones (of sizes up to 30). 
Thus, we treat this as a special reduction rule that applies a new instance of the solver
to every connected component but the largest one, reduces them with the best solution found,
   and continues solving the largest connected component. 
\item[solving subcubic instances] As proven by~\cite{fvs:3.83k}, if every vertex $v \in V(G) \setminus U$ is of degree at most three, then the corresponding
instance is polynomial-time solvable.
Kociumaka and Pilipczuk~\cite{KociumakaP} provided a simpler proof of this result
via a reduction to the matroid parity problem in graphic matroids, which proceeds as follows.
First, apply the known reduction rules that reduce the problem to the case when every $v \in V(G) \setminus U$ is of degree exactly three.
Second, subdivide every edge $uv \in E(G-U)$ with a new vertex $x_{uv} \in U$, so that $V(G) \setminus U$ is an independent set.
Third, for every $v \in V(G) \setminus U$, pick two out of the three edges incident to $v$ as a pair $P_v$.
Let $Q = E(G) \setminus \bigcup_{v \in V(G) \setminus U} P_v$ be the remaining edges; note that every $v \in V(G) \setminus U$ is of degree exactly $1$ in
the graph $(V(G), Q)$. Then, it is easy to observe that the \textsc{Feedback Vertex Set} problem is equivalent to picking a set $Y \subseteq V(G) \setminus U$
of maximum cardinality such that $(V(G), Q \cup \bigcup_{v \in Y} P_v)$ is a forest.
This condition is the same as requiring $Y$ to be a maximum matching in the graphic matroid of the graph $G/Q$ ($G$ with edges of $Q$ contracted) with pairs
$(P_v)_{v \in V(G) \setminus U}$. \\
In some of our algorithms, we use the approach of~\cite{KociumakaP} to solve such instances.
As the underlying solver to the matroid matching problem, we use
the augmenting path algorithm of Gabow and Stallmann~\cite{GabowS86}.
\item[solution lower bound] Consider an instance $(G,U,k)$ and let
$v_1,v_2,\ldots$ be the  vertices of $V(G) \setminus U$ in the nonincreasing
order of degrees in $G$.
If $(G,U,k)$ admits a solution $X$ of size $j$, then 
$G-X$ has at least $|E(G)|-\sum_{i=1}^j \mathrm{deg}_G(v_i)$ edges.
On the other hand, if $G-X$ is a forest, it has less than $|V(G)|-j$ edges.
Consequently, we can safely stop if for all $0 \leq j \leq k$ we have that
$$|E(G)|-\sum_{i=1}^j \mathrm{deg}_G(v_i) \geq |V(G)|-j.$$
The above prunning strategy has been used
in the entry of Imanishi and Iwata~\cite{IwataGithub}.
\end{description}
Unless otherwise noted, all our implementations split instances
into connected components. We also compare a number of selected approaches
without this preprocessing step to see its impact on performance.

The algorithm of Cao~\cite{Cao} uses all aforementioned simple reduction rules
as well as the solver of subcubic instances. 
On a branching step, it simply chooses the vertex of highest degree.
As shown by Cao, such an algorithm has running time bound $8^k n^{\Oh(1)}$.
We also test a variant of the algorithm of Cao that first
branches on vertices incident with double edges,
a variant that does not use the subcubic instance solver,
  and a variant that prunes the search tree via the aforementioned lower bound.

\subsection{Approximation and iterative compression}\label{ss:apx}

The algorithms of~\cite{fvs:5k,fvs:3.83k,KociumakaP} operate in the framework
of iterative compression. That is, their central subroutine
solves a seemingly simpler problem, where additionally a slightly too large
solution $Y$ is given, and the algorithm first branches on the vertices
of $Y$ (putting each $y \in Y$ into the solution or into set $U$).
Since at the beginning $G-Y$ is a forest, and every vertex of $Y$ is either deleted or put into $U$,
      we obtain the property that $G-U$ is a forest.
This greatly helps in the analysis.

In the literature, the set $Y$ is traditionally taken from the iterative 
compression step. One picks an order $V(G) = \{v_1,v_2,\ldots,v_n\}$
and solves iteratively \textsc{Feedback Vertex Set} on graphs
$G_i = G[\{v_1,v_2,\ldots,v_i\}]$. Given a solution $X_{i-1}$ to $G_{i-1}$,
one can set $Y = X_{i-1} \cup \{v_i\}$ for $G_i$.

However, such an approach leads to multiple invocation of the same algorithm,
and a substantial multiplicative overhead in the running time bound.
In our algorithms, we instead find $Y$ via a simple heuristic:
reduce the graph via simple reductions as long as possible and, when impossible,
delete the vertex of highest degree.
In Section~\ref{sec:results} we discuss the performance of this heuristic
on our test data.

Our branching framework keeps a queue of \texttt{branching hints}
and, if nonempty, the algorithm always branches on a vertex from the queue.
For algorithms based on iterative compression,
the queue is initiated by
an approximate solution found by our heuristic.
This corresponds to passing the set $Y$
to the algorithms based on iterative compression, but allows to reduce
some of the vertices of the set $Y$ by reductions after a number of branching steps.
If an algorithm does not use iterative compression, the queue is empty through
the entire run of the algorithm.

The algorithm of~\cite{fvs:5k} implements the iterative compression framework
and branches on a vertex of $V(G)\setminus U$ that is incident to maximum
number of edges leading to $U$.
As shown in~\cite{fvs:5k}, such an algorithm has $5^k n^{\Oh(1)}$ time
bound guarantee.

The algorithm of~\cite{KociumakaP} is arguably a simplification of
the arguments of~\cite{fvs:3.83k}, so we implement only the first one. 
It modifies the algorithm of~\cite{fvs:5k} in the following way:
\begin{itemize}
\item It leaves alone vertices $v \in V(G) \setminus U$ that are of degree $3$
and all their incident edges lead to $U$ (such vertices are 
called henceforth \emph{tents}). The crux is that if every vertex $v \in V(G) \setminus U$ is a tent,
then we can apply the polynomial-time algorithm of~\cite{fvs:3.83k,KociumakaP} to the instance.
In other words, tents form a ``polynomial-time solvable'' part of the instance.
\item Given a vertex $v \in V(G) \setminus U$ of degree $3$ with exactly one
neighbor $u$ in $V(G) \setminus U$ (and other $2$ neighbors in $U$), it proceeds as follows:
\begin{itemize}
\item subdivide the edge $uv$ with a new vertex $w \in U$; note that this does not change the set of feasible solutions to the \textsc{Feedback Vertex Set} problem;
\item marks $w$ irreducible for the reduction suppressing degree-2 vertices.
\end{itemize}
Note that this operation turns $v$ into a tent, while reducing the number of vertices of $V(G) \setminus U$ that are not tents.
\item It applies the solver for subcubic instances if every vertex of $V(G) \setminus U$ is a tent.
\end{itemize}
As shown in~\cite{KociumakaP}, such an algorithm has $3.62^k n^{\Oh(1)}$ time
bound guarantee.

Inspired by the methods of choice of branching pivots of
the algorithms~\cite{fvs:5k,fvs:3.83k,KociumakaP}, we also test
a variant of Cao's algorithm where the choice of the branching pivot 
is as in~\cite{fvs:5k}: vertex with maximum number of neighbors
in $U$ (but, contrary to~\cite{fvs:5k}, no iterative compression).

Additionally, we check how much the algorithms can be sped up by adding
prunning via the aforementioned lower bound
and if the Cao's algorithm can benefit from 
the use of iterative compression.

\subsection{Branching based on half-integral relaxation}

Iwata, Wahlstr\"{o}m, and Yoshida showed a generic approach to
numerous transversal problems via half-integral relaxations~\cite{IwataWY16}.
They are all based on the following principle: a half-integral relaxation
of a variant of the problem is defined and shown to be polynomial-time solvable.
Furthermore, the solution to the half-integral relaxation has some
persistency property: it either indicates some greedy choice for
the integral problem, or indicates a good branching pivot.
In this approach, the time needed to find a solution of the half-integral
relaxation is critical.

The algorithms of~\cite{IwataWY16} run in linear time
for edge deletion problems, but unfortunately for vertex-deletion problems
the subroutine that finds the half-integral relaxation requires
solving linear programs. 
The main contribution of Iwata~\cite{Iwata17} (implemented in the PACE
2016 entry by Iminishi and Iwata~\cite{IwataGithub})
is a combinatorial linear-time solver for the half-integral relaxation 
in the special case of \textsc{Feedback Vertex Set}.
We reimplement this solver in our branching framework.

Iwata~\cite{Iwata17} also observed that the half-integral relaxation
can be used to find a polynomial kernel for the problem, improving
the previous seminal kernel of Thomass\'{e}~\cite{fvs:quadratic-kernel}.
Apart from the reduction rules mentioned before, this kernel employs another
involved reduction rule, applicable on vertices of degree more than $2k$.
All our implementations based on a half-integral relaxation implement
this preprocessing step as well.

The half-integral relaxation of~\cite{IwataWY16,Iwata17} does not solve
\textsc{Feedback Vertex Set} directly, but rather given an undeletable
vertex $u \in U$, tries to separate an acyclic connected component
(i.e., a tree) containing $u$ from the rest of the graph.
On high level, the branching strategy of the algorithm is as follows.
If $U = \emptyset$, we branch on any vertex; we choose
one of highest degree here for efficiency.
Otherwise, we use a vertex $u \in U$ as a root for the half-integral relaxation.
The persistence properties of the relaxation ensure that we
can perform a greedy step unless the relaxation put values $0.5$
on all neighbors of $u$. If this is the case, we branch on a neighbor
of $u$; we choose highest-degree neighbor here for efficiency.
Note that once a tree component with $u$ gets separated, simple reduction
rules delete it from the graph.

Since the kernelization routine of~\cite{Iwata17} is computationally
expensive, it is not obvious if one should apply it at every step.
We experiment with two variants: when we run the kernelization step
at every step, or only at steps with $U = \emptyset$.
Furthermore, we also check how much prunning with the lower bound heuristic
or solver of subcubic instances helps.

\subsection{Using a generic Integer Linear Program solver}

As mentioned in the introduction section, we can create an integer linear programming 
formulation of our problem in two ways. For purpose of the tests, we have chosen the
cycle-based implementation created by organizers of PACE challange~\cite{PACE2016tests}, 
 which as a linear program looks like:

\begin{align*}
\mathrm{minimize}\qquad & x_1 + x_2 + ... + x_{|V(G)|} \\
\mathrm{subject\ to}\qquad & x_v \in \{0, 1\} & \forall v\in V(G) \\
 & x_{c_1} + x_{c_2} + ... + x_{|c_{|C|}|} \geq 1 & \forall \mathrm{cycle}\ C=(x_{c_1}, ..., x_{|c_{|C|}|})
\end{align*}

One may notice that the number of constraints on the cycles can be exponential in the size of the graph. For example in a clique on $n$ vertices, there are roughly $n!$ cycles.
For this reason we cannot just add all of the constraints at once --- we want to make the graph smaller first and than add the constraints dynamically.

The implementation first applies the reductions $2.-5.$, mentioned in the beggining of the chapter, to the input graph.
Next it removes the bridges between connected components of the reduced graphs.
Afterwards it processes every component of the graph separately in the steps described below.
\begin{itemize}
	\item Put a heuristic solution as the starting points of the ILP ($v \in X \Rightarrow x_v = 1$).
  Here we use a slight modification of the heuristic: we take shortest cycles one by one, find the highest degree vertex $v$ in them, and add $v$ to $X$.
  Repeat this as long as there are any cycles in $G$.
	\item Find a set of cycles of shortest length (run BFS from every vertex) and add them to our constraint pool.
	All other constraints are lazy, which means at first the solver finds a feasible solution without using them. 
	Later if evaluation turns out to violate them, they are being added and the solution is recalculated.
	\item In the callbacks, we check whether the solver returned a set $X$ that actually removes all the cycles in $G$. 
	If it is the case, then we have found a solution for our problem. 
	Otherwise, we return to the point of adding new shortest-cycle constraints. 
	In order to only add new cycles, we delete from the graph vertices, 
	which were part of the latest removal set $X$, and then compute shortest cycles.
\end{itemize}
During our tests we have used Gurobi $8.0$ optimizer.

\section{Experiment setup}\label{sec:setup}

\subsection{Hardware and code}

The experiments have been performed on a cluster of
16 computers at the Institute of Informatics, University of Warsaw.
Each machine was equipped with Intel Xeon E3-1240v6 3.70GHz processor
and 16 GB RAM.
All machines shared the same NFS drive. Since the size of the inputs and
outputs to the programs is small, the network communication
was negligible during the process.

The code has been writen in C++ and is available at~\cite{our-repo} or at project's
webpage~\cite{fnp-webpage}.

\subsection{Test cases}

As discussed in the introduction, we repeat the setup of the
PACE 2016 challenge~\cite{PACE2016}.
At PACE 2016, the organizers gathered 230 graphs from different sources~\cite{PACE2016tests}.
A subset of 100 of them has been made public prior to the competition
deadline, and the final evaluation has been made on the hidden
130 instances.
We run every tested algorithm on each of the 230 instances with 30 minutes
timeout.

Out of the test instances, we gathered two subsets to compare actual
running times of the algorithm. The first set, $A$, consists
of test cases solved by all algorithms, but with substantial running time
of some of them. The second set, $B$, is defined similarly, but with
regards only to the algorithms that use prunning via the lower bound.
The third set, $C$, is a subset of $A$ and $B$ that has been solved by the ILP solver.

More precisely, set $A$ consists of the following 14 tests:

\noindent\begin{tabular}{lllllll}
\texttt{hidden\_001} & \texttt{hidden\_007} & \texttt{hidden\_012} & \texttt{hidden\_056} & \texttt{hidden\_065} & \texttt{hidden\_083} & \texttt{hidden\_099} \\
\texttt{hidden\_106} & \texttt{public\_011} & \texttt{public\_014} &
\texttt{public\_037} & \texttt{public\_069} & \texttt{public\_076} & \texttt{public\_086} \\
\end{tabular}

The set $B$ consists of the following 7 tests:

\noindent\begin{tabular}{lllllll}
\texttt{hidden\_022} & \texttt{hidden\_041} & \texttt{hidden\_068} & \texttt{hidden\_088} & \texttt{public\_035} &
\texttt{public\_066} & \texttt{public\_067} \\
\end{tabular}

The set $C$ consists of the following 18 tests:

\noindent\begin{tabular}{lllllll}
\texttt{hidden\_001} & \texttt{hidden\_007} & \texttt{hidden\_012} & \texttt{hidden\_022} & \texttt{hidden\_056} & \texttt{hidden\_065} & \texttt{hidden\_068} \\
\texttt{hidden\_083} & \texttt{hidden\_088} & \texttt{hidden\_099} & \texttt{hidden\_106} &
\texttt{public\_011} & \texttt{public\_014} & \texttt{public\_035} \\
\texttt{public\_069} & \texttt{public\_076} & \texttt{public\_086} & \texttt{public\_067} &&& \\
\end{tabular}

We also gather sizes of approximate solution found by our heuristic
and compare it with the optimal size found by the algorithm.
 
\subsection{Reduction rules measures}\label{subsec:redMes}

In order to fully compare the branching algorithms, 
we have decided to measure how big impact do actually the reduction rules have on 
performance of each algorithm. 
For this purpose we introduce the following measures.
We computed for every test and for the top $3$ algorithms in our benchmark.
\begin{description}
	\item[initial reductions] How well the reductions work before the first use of branching. 
	There are both numbers of how many vertices/edges were reduced initially and
	how big part of the whole graph it was (in percents).

  Here most of the algorithms use the same set of initial rules, except for the
  algorithm by Imanishi and Iwata that uses an extra kernelization step it uses.
	\item[average over recursive subcalls] Measurements of how big impact do the reductions have during the runtime of branching.
	We count an average number of vertices/edges reduced during one step of branching
  (averaged over all recursive calls to the branching subroutine).
	In this group we have also counted how big impact did reduction steps have only in branches where the number of vertices in part of graph remaining to solve is between $20$ and $40$.
	\item[connected components] In this part we have calculated how big are the connected components separated from the largest connected component of the input graph by our "split into connected components" routine. 
\end{description}

One may find exact results of above measurements for each of the tests in our repository~\cite{our-repo}.

We analyze performance of reductions on each algorithm by grouping tests with following methods.
\begin{itemize}
	\item Average over all tests finished by all of the three algorithms.
	\item Score function is calculated in two ways. In the comparison of the initial reductions, since we are comparing only two algorithms, the score is the number of tests on which reductions in one solution have strictly better results than in the others.
  In the other comparisons, since they compare three approaches, for every test 
	we divide $1$ point between all solutions with best result on the test. 
  We report the percentage of total points scored by each algorithm.
\end{itemize}

While the measurement of initial reductions is a solid comparison of the simple reduction rules
with the addition of the sophisticated kernelization rule, 
     the remaining measurements are just our quite crude attempts at measuring how well does the algorithm perform reductions of the graph during the whole branching routine, and should be treated with a grain of salt.

\section{Results}\label{sec:results}

A full table with running times of each program
at each test can be found at the repository~\cite{our-repo}.

\subsection{Performance of the approximation heuristic}

We compared the performance of the approximation heuristic discussed in Section~\ref{ss:apx}
(i.e., greedily take the highest-degree vertex after applying the simple reduction rules)
with the size of the optimum solution that 
was known to us on 127 instances.
The results are in Table~\ref{tb:apx}.
On only $8$ instances, the approximate solution was more than one vertex
larger than the optimum one. 
Consequently, the approximate solution can serve well as the basis
for iterative compression.
\begin{table}[htb]
\begin{center}
\begin{tabular}{l|cccc|c}
difference approximation minus optimum & 0 & 1 & 2 & $>2$ & $>10\% \cdot \textrm{optimum}$ \\\hline
number of instances & 89 & 30 & 4 & 4 & 5\\
\end{tabular}
\caption{Comparison of the size of the approximate and exact solution found
  on 127 instances.}\label{tb:apx} 
  \end{center}
\end{table}

\subsection{Comparison}

We have run 22 different algorithms on the whole test data.
A CSV file with full results is available at the repository~\cite{our-repo}.
Table~\ref{tb:cmp} contains aggregated values: number of solved test instances within the time limit (30 minutes per instance)
and the total running time on sets $A$, $B$, and $C$.
Please see the caption of Table~\ref{tb:cmp} for description of the notation used in the table.

\begin{table}[htb]
\begin{center}
\begin{tabular}{l|llll|lll|l|l|l}
\multirow{2}{*}{algorithm}  & \multicolumn{4}{c}{optimizations} & \multicolumn{3}{c}{solved instances} & \multicolumn{3}{c}{total time (MM:SS.ms)} \\
  & CC & deg3 & LB & IC & all & public & hidden & set $A$ & set $B$ & set $C$ \\\hline
       Cao &   & + &   &   &        100 &         48 &         52 &   35:17.21 &          - &         - \\
     CFLLV &   &   &   & + &         91 &         44 &         47 &   35:16.06 &          - &         - \\
        KP &   & + &   & + &        101 &         49 &         52 &   36:22.48 &          - &         - \\
       Cao & + & + &   &   &        101 &         48 &         53 &   37:31.83 &          - &         - \\
     CFLLV & + &   &   & + &         91 &         44 &         47 &   31:23.63 &          - &         - \\
        KP & + & + &   & + &        101 &         49 &         52 &   32:00.49 &          - &         - \\
       Cao & + &   &   &   &         91 &         43 &         48 &   38:00.78 &          - &         - \\\hline
        II & + &   &   &   &         92 &         46 &         46 &   34:51.17 &          - &         - \\
II/kernel  & + &   &   &   &         90 &         45 &         45 &   75:11.16 &          - &         - \\\hline
       Cao & + & + & + &   &        123 &         62 &         61 &    0:19.28 &   10:38.88 &    1:54.11 \\
Cao/double & + & + & + &   &        122 &         62 &         60 &    3:22.52 &   21:31.54 &   15:57.95 \\
Cao/undel  & + & + & + &   &        118 &         58 &         60 &    0:35.67 &    8:39.05 &    8:12.94 \\
       Cao & + & + & + & + &        123 &         62 &         61 &    0:19.99 &   10:37.88 &    1:53.48 \\
Cao/double & + & + & + & + &        122 &         62 &         60 &    3:13.71 &   14:31.66 &    8:45.33 \\
Cao/undel  & + & + & + & + &        118 &         58 &         60 &    0:36.88 &    8:12.80 &    7:47.82 \\
     CFLLV & + &   & + & + &        117 &         57 &         60 &    0:37.61 &    8:12.88 &    7:47.68 \\
        KP & + & + & + & + &        118 &         58 &         60 &    0:38.95 &    8:28.46 &    8:03.57 \\
       Cao & + &   & + &   &        122 &         61 &         61 &    0:22.94 &   10:28.32 &    1:58.57 \\\hline
        II & + &   & + &   &        117 &         58 &         59 &    1:36.79 &   25:31.95 &   24:29.82 \\
II/kernel  & + &   & + &   &        109 &         55 &         54 &    7:24.42 &          - &          - \\
        II & + & + & + &   &        118 &         59 &         59 &    1:37.08 &   25:33.76 &   24:33.01 \\

\hline
       ILP &   &   &   &   &        140 &         62 &         78 &          - &          - &    3:46.37 \\\hline

\end{tabular}
\caption{Comparison of different algorithms.
The first column indicates the base of the algorithm: Cao's~\cite{Cao}, CFLLV for Chen et al.~\cite{fvs:5k}, KP for Kociumaka and Pilipczuk~\cite{KociumakaP}, II for Imanishi and Iwata~\cite{IwataGithub,Iwata17} and ILP for Integer Linear Programming. II/kernel stands for the algorithm of~\cite{IwataGithub,Iwata17} that runs
  the kernelization step more often, namely at every branching step. 
  Cao/double stands for the algorithm of~\cite{Cao} that first branches on double edges.
  Cao/undel stands for the algorithm of~\cite{Cao} that chooses branching pivot with regards to maximum degree to undeletable vertices.
  In the optimizations columns, CC stands for splitting into connected components, deg3 for the use of solver of subcubic instances,
LB for the use of prunning with the lower bound, and IC stands for iterative compression.}
  \label{tb:cmp}
\end{center}
\end{table}

The first nine algorithms do not use prunning via the lower bounding technique, and the first three do not use
splitting into connected components. 
They are mostly meant to compare basic approaches.

Without the lower bound prunning, the best approaches are Cao's~\cite{Cao} and Kociumaka-Pilipczuk~\cite{KociumakaP},
        and they seem to be rather incomparable.
The other algorithm based on iterative compression of Chen et al.~\cite{fvs:5k} is clearly outperformed by the other two,
and the same holds for the branching algorithm based on half-integral relaxation~\cite{IwataGithub,Iwata17}.

The first three rows differ from rows 4-6 by the usage of splitting into connected components. They show that the effect
of this improvement is small, and even hurt a bit Cao's algorithm~\cite{Cao}.

The 7th row treats Cao's algorithm~\cite{Cao} without the solver of the subcubic instances. It indicates that this solver
is essential for the performance of Cao's algorithm.

The 9th row treats the Imanishi-Iwata algorithm~\cite{IwataGithub,Iwata17} with more often application of the kernelization step,
namely at every branching step (not only at the ones with $U = \emptyset$). It shows that the step is too expensive to execute
it that often.

Let us now discuss the algorithms with the lower bound prunning step.
First, the results show that the prunning step greatly improves performance for all algorithms.

With regards to the variants of Cao's algorithm~\cite{Cao}, the results show that any mutation of the branching pivot rule
here is undesirable. Also, adding the iterative compression step does not seem to have any particular impact on the performace.
Interestingly, the negative effect of dropping the solver of the subcubic instances mostly disapper if one
adds the prunning step.

For the branching algorithm based on the half-integral relaxation~\cite{IwataGithub,Iwata17}, the last three rows
of Table~\ref{tb:cmp} covering II again confirm the corollary that more often application of the kernelization step is undesirable.
There is little difference with addition of the solver of subcubic instances (the main difference comes from the fact that the test set
contains one huge cubic graph, \texttt{public\_84}, which is not amenable to any branching technique we tested).

Finally, in our experiments the best variant of Cao's algorithm~\cite{Cao} with the prunning slightly outperforms the
best variant of the branching algorithm based on half-integral relaxations~\cite{IwataGithub,Iwata17}.
However, the difference (5 tests more and roughly $3\times$ speed-up on sets $A$ and $B$)
  is not big enough to decisively conjecture its advantage.

First, it is possible that a $3\times$ speed-up can be gained by low-level optimizations of the 
solver of the half-integral relaxations. Arguably, Cao's algorithm~\cite{Cao} is much simpler and thus easier to optimize.
Second, it may be also an artifact of the chosen test data: there are 5 tests in the data set
that were solved by the penultimate algorithm in Table~\ref{tb:cmp}, but not by the 10th (and 10 tests vice-versa).
That is, there are types of instances solved significantly faster by one of the approaches but not by another.

We conclude with a remark about comparison with PACE'16 results~\cite{PACE2016}, where the entry of Imanishi and Iwata~\cite{IwataGithub} solved 84 hidden instances
while the entry of the second author~\cite{PilipczukPACE}, based on Cao's algorithm~\cite{Cao}, solved 66. 
While this data seemingly stands in contradiction with the results in Table~\ref{tb:cmp}, one should note that the two entries differ significantly
in other internals. Most importantly, the first entry used prunning with the lower bound, while the second one did not.
Other difference includes: removal of bridges in the first entry vs splitting into 2-connected components in the second,
and the use of bounded treewidth subroutine in the second entry.
In other words, our results indicate that the big difference in the performance of the first two entries at PACE 2016 were mainly caused by the difference
in preprocessing routines and prunning heuristics (and possibly low-level optimizations) rather than in the underlying base branching algorithm.

The last algorithm in the table is the solution based on integer linear programming. 
It is the solver that the organizers used while preparing the contest.
We may notice that at the public tests set it solves pretty much the same amount of tests ($62$) as the best branching algorithms. 
On the other hand, the difference can be noticed on the set of hidden tests, where it solves $17$ tests more than any other solution we have tested. 
We expected ILP solution to greatly outperform all of the solutions on both sets of tests, so the results can be found a little bit surprising.

\subsection{Impact of the reduction rules}

In this section we will analyse the impact of reduction rules on each of the branching algorithms that achieved the best results in our tests, i.e. Cao, KP, and II.
The measures were introduced in Section~\ref{subsec:redMes}.
In what follows, each average over tests is taken over all tests solved by all three solvers taken into account.

\begin{table}[h]
\begin{center}
\begin{tabular}{c|cc|c}
    & Cao & KP & II \\
    \hline
  Average $\Delta n$ \% & \multicolumn{2}{c|}{49.24\%} & 53.31\%   \\
  $\Delta n$ score & \multicolumn{2}{c|}{0} & 14   \\
  Average $\Delta m$ \% & \multicolumn{2}{c|}{42.32\%} & 46.94\%   \\
  $\Delta m$ score & \multicolumn{2}{c|}{4} & 16   \\
  \hline
\end{tabular}
\end{center}
\caption{Initial reductions: the impact of reduction rules before start of a branching.
$\Delta n$ stands for the decrease in the number of vertices, $\Delta m$ for the decrease in 
the number of edges. Score indicates the number of tests where one algorithm
strictly outperformed the second one.}
\end{table}

\paragraph{Initial reductions.}
On average, reduction rules get rid of almost $50\%$ of the graph even before the start of branching. This shows a great impact of preprocessing on further performance of the algorithms.

From the $\Delta n$ score, we may notice that II has at least as high result as the other algorithms in $\Delta n \%$ on every test. 
Recall that all algorithms except II use the same set of reductions, while the II algorithm uses
an extra sophisticated kernelization step for vertices of large degree.
The II algorithm achieves better performance on $14$ tests (in terms of the number of reduced vertices) due to the extra kernelization step.

Note that II looses in $4$ tests when calculating $\Delta m$ score. 
The reason for this is pretty simple: the kernelization step, while changing structure of the graph is not only deleting vertices and edges, but also adding some edges to preserve the potential solution unchanged. In those tests it happened that more edges were added than deleted.

\begin{table}[h]
\begin{center}
\begin{tabular}{c|c|c|c}
    & Cao & KP & II \\
    \hline
    AVG -N score & 56.91\% & 20.42\% & 22.67\%    \\
    AVG -M score & 54.20\% & 16.82\% & 28.98\%    \\
    AVG20-40 -N score & 52.86\% & 17.86\% & 29.29\%    \\
    AVG20-40 -M score & 35.71\% & 20.71\% & 43.57\%    \\
    \hline
\end{tabular}
\end{center}
\caption{Average report of reduction rules performance in the runtime of each branching.
For each test, we distribute one point between the best performing algorithms,
and then present the percentage of all points got by each of the algorithms.}
\end{table}

\paragraph{Average over recursive calls.}
As mentioned in the previous section, we present the number of reduced vertices and edges
averaged over all recursive calls of the branching algorithm and over all calls that 
are applied to graphs with at least 20 and at most 40 vertices.

The results of the first part clearly show that Cao reduces most vertices/edges on average during one step of branching. 
The explanation for this result is the way we choose vertex for branching in implementation of Cao. 
Choosing vertex of maximum degree (with most edges) for branching favors creation of new reducible vertices. 

Even though relative order of algorithms for AVG20-40 vertices score hold, 
one may wonder why that is not true for score on AVG20-40 edges reductions. 
It might happen that Cao's average reduced edges number for instances above $40$ is close to some other algorithm (both of the others gain for instances in $20$-$40$). 
Branching on highest degree vertices when the instance is still big, makes us get rid of huge part of the edges, i.e., the number of edges Cao has left to reduce becomes smaller much faster than for the other algorithms. 
For this reason, when we get to medium sized instances it may turn out that some other algorithm that had close average number of reduced edges has it higher now and Cao loses point from its score on that test. 

\begin{table}[h]
\begin{center}
\begin{tabular}{c|c|c|c}
    & Cao & KP & II \\
    \hline
    Average total separated vertices & 3201.93 & 20810.23 & 18539.40    \\
    Score total separated vertices & 29.43\% & 35.29\% & 35.29\%    \\
    \hline
\end{tabular}
\end{center}
\caption{Connected components separation performance results.
For each test, we distribute one point between the best performing algorithms,
and then present as the score the percentage of all points got by each of the algorithms.}
\end{table}

\paragraph{Connected components.} In the table covering connected components we can see that on average the KP and II algorithms seem to separate the biggest number of vertices by cutting out connected components from the graph.
If we look at the results for separate tests, it seems that on average sizes of connected components cut out from graphs are small and similar for all of the algorithms. From this we could infer that KP and II cut out connected components much more often than Cao. 
It stands opposite to the intuition of how the II works: 
it tries to solve whole connected components on its own during the branching step. 
The reason for this abnormality in this measure is $1$ test on which both KP and II did a huge number of separations (higher than the sum of all the other tests together), while Cao did almost none. 

For this reason we find the measure of score much more meaningful this time. It says that every algorithm won on almost the same amount of tests and the impact of connected components separation on KP and II is not that much bigger than on Cao.

\subsection{Comparison of solved tests}

We have noticed that the $5$ test difference between Cao and II algorithms does not
mean straightforward that Cao solved $5$ tests that II did not.
It actually is a little bit different -- II solved $5$ tests that Cao did not and 
there were $10$ tests the other way.
We have decided to note down the tests on which they differ and 
analyze them in order to find some structural property that makes one algorithm perform better
on one set than the other.
Below you can find the list of tests solved only by one of the algorithms.

List of tests that Cao solved and II did not:

\noindent\begin{tabular}{lllll}
\texttt{hidden\_023} & \texttt{hidden\_035} & \texttt{hidden\_064} & \texttt{hidden\_066} & \texttt{hidden\_109} \\
\texttt{public\_032} & \texttt{public\_034} &
\texttt{public\_073} & \texttt{public\_080} & \texttt{public\_087} \\
\end{tabular}

List of tests that II solved and Cao did not:

\noindent\begin{tabular}{lllll}
\texttt{hidden\_071} & \texttt{hidden\_079} & \texttt{hidden\_080} & \texttt{public\_023} & \texttt{public\_043} \\
\end{tabular}

While analyzing the tests, we have first applied the basic reductions on the graphs and than generated statistics like number of vertices, number of edges, number of double edges and the distribution of degrees in the graph. 
You may find the statistics in our repository.
Afterwards we used \texttt{gephi} to draw and look at the graphs. 

We have found out that in the tests that Cao solved there is hardly any regularity to be found (with one exception of test $\texttt{hidden\_109}$, which is an almost-regular graph with low degrees). 
In these tests there usually are some vertices of high degree and many low degree vertices. 
Such structure of the test makes Cao algorithm work fast: it branches on the vertices of highest degree first. Thanks to that we quickly get rid of huge part of the graph: having decided whether such a vertex is in solution $X$ or in safe $U$ set, degrees of many vertices may decrease. 
Now having many vertices of low degrees helps:
the decrease in degrees affects a lot of them, which at the same time makes the basic reductions quickly become applicable again. 
At the same time on such tests II is much computationally-heavier and is unable find any structure that it could solve quickly, which is  reason for this difference.

On the other hand looking at the tests solved by II and not by Cao we can see a lot of regularity.
Here, we have regular graphs with degrees in range from 10 to 16. 
Because of it, Cao is completely ineffective on these instances. It has to branch through a lot of vertices to achieve anything, which is almost impossible with limited time it is given. 
II seems to work on such graphs more efficiently. It generates less branches and gets rid of such grid-like structures faster.

\section{Conclusions}\label{sec:conc}

We have conducted a thorough experimental evaluation of various
parameterized algorithms for the \textsc{Feedback Vertex Set} problem,
following the setup of Parameterized Algorithms and Computational Experiments
Challenge from 2016~\cite{PACE2016}.
Our results does not confirm greater advantage of the approach
based on half-integral relaxations that was suggested by PACE'16 results,
but rather suggest that lower bounding techniques and low-level
optimizations were decisive at PACE'16.

On the other hand, the approach via half-integral relaxation turned
out to be almost on par with the best variant of Cao's algorithm~\cite{Cao}.
This still indicates big potential in this approach, in particular
in the light of the recent (theoretical) generalization of
this approach to other problems~\cite{IwataYY17}. 
Experimental evaluation of this approach for other problems such as
\textsc{Multiway Cut} and \textsc{Odd Cycle Transversal}
is a topic for future work.

\bibliographystyle{abbrv}
\bibliography{refs}

\begin{thebibliography}{10}

\bibitem{fnp-webpage}
Recent trends in kernelization: theory and experimental evaluation --- project
  website.
\newblock 2018.
\newblock \texttt{http://kernelization-experiments.mimuw.edu.pl}.

\bibitem{AkibaI16}
T.~Akiba and Y.~Iwata.
\newblock Branch-and-reduce exponential/fpt algorithms in practice: {A} case
  study of vertex cover.
\newblock {\em Theor. Comput. Sci.}, 609:211--225, 2016.

\bibitem{fvs:4krand}
A.~Becker, R.~Bar-Yehuda, and D.~Geiger.
\newblock Randomized algorithms for the loop cutset problem.
\newblock {\em J. Artif. Intell. Res. (JAIR)}, 12:219--234, 2000.

\bibitem{fvs1}
H.~L. Bodlaender.
\newblock On disjoint cycles.
\newblock {\em Int. J. Found. Comput. Sci.}, 5(1):59--68, 1994.

\bibitem{fvs:kernel2}
H.~L. Bodlaender and T.~C. van Dijk.
\newblock A cubic kernel for feedback vertex set and loop cutset.
\newblock {\em Theory Comput. Syst.}, 46(3):566--597, 2010.

\bibitem{fvs:kernel1}
K.~Burrage, V.~Estivill-Castro, M.~R. Fellows, M.~A. Langston, S.~Mac, and
  F.~A. Rosamond.
\newblock The undirected feedback vertex set problem has a $poly(k)$ kernel.
\newblock In H.~L. Bodlaender and M.~A. Langston, editors, {\em IWPEC}, volume
  4169 of {\em Lecture Notes in Computer Science}, pages 192--202. Springer,
  2006.

\bibitem{Cao}
Y.~Cao.
\newblock A naive algorithm for feedback vertex set.
\newblock In R.~Seidel, editor, {\em 1st Symposium on Simplicity in Algorithms,
  {SOSA} 2018, January 7-10, 2018, New Orleans, LA, {USA}}, volume~61 of {\em
  {OASICS}}, pages 1:1--1:9. Schloss Dagstuhl - Leibniz-Zentrum fuer
  Informatik, 2018.

\bibitem{fvs:3.83k}
Y.~Cao, J.~Chen, and Y.~Liu.
\newblock On feedback vertex set new measure and new structures.
\newblock In H.~Kaplan, editor, {\em SWAT}, volume 6139 of {\em Lecture Notes
  in Computer Science}, pages 93--104. Springer, 2010.

\bibitem{fvs:5k}
J.~Chen, F.~V. Fomin, Y.~Liu, S.~Lu, and Y.~Villanger.
\newblock Improved algorithms for feedback vertex set problems.
\newblock {\em J. Comput. Syst. Sci.}, 74(7):1188--1198, 2008.

\bibitem{fvs:3k}
M.~Cygan, J.~Nederlof, M.~Pilipczuk, M.~Pilipczuk, J.~M.~M. van Rooij, and
  J.~O. Wojtaszczyk.
\newblock Solving connectivity problems parameterized by treewidth in single
  exponential time.
\newblock In R.~Ostrovsky, editor, {\em FOCS}, pages 150--159. IEEE, 2011.

\bibitem{fvs7}
F.~K. H.~A. Dehne, M.~R. Fellows, M.~A. Langston, F.~A. Rosamond, and
  K.~Stevens.
\newblock An $\uppercase{O}(2^{\uppercase{o}(k)}) n^3$ {FPT} algorithm for the
  undirected feedback vertex set problem.
\newblock {\em Theory Comput. Syst.}, 41(3):479--492, 2007.

\bibitem{PACE2016}
H.~Dell, T.~Husfeldt, B.~M.~P. Jansen, P.~Kaski, C.~Komusiewicz, and F.~A.
  Rosamond.
\newblock The {F}irst {P}arameterized {A}lgorithms and {C}omputational
  {E}xperiments {C}hallenge.
\newblock In J.~Guo and D.~Hermelin, editors, {\em 11th International Symposium
  on Parameterized and Exact Computation, {IPEC} 2016, August 24-26, 2016,
  Aarhus, Denmark}, volume~63 of {\em LIPIcs}, pages 30:1--30:9. Schloss
  Dagstuhl - Leibniz-Zentrum fuer Informatik, 2016.

\bibitem{PACE2017}
H.~Dell, C.~Komusiewicz, N.~Talmon, and M.~Weller.
\newblock {The {PACE} 2017 {P}arameterized {A}lgorithms and {C}omputational
  {E}xperiments {C}hallenge: {T}he {S}econd {I}teration}.
\newblock In D.~Lokshtanov and N.~Nishimura, editors, {\em 12th International
  Symposium on Parameterized and Exact Computation (IPEC 2017)}, Leibniz
  International Proceedings in Informatics (LIPIcs), pages 30:1--30:12,
  Dagstuhl, Germany, 2018. Schloss Dagstuhl--Leibniz-Zentrum fuer Informatik.

\bibitem{fvs2}
R.~G. Downey and M.~R. Fellows.
\newblock Fixed parameter tractability and completeness.
\newblock In {\em Complexity Theory: Current Research}, pages 191--225, 1992.

\bibitem{fvs3}
R.~G. Downey and M.~R. Fellows.
\newblock {\em Parameterized Complexity}.
\newblock Springer, 1999.

\bibitem{PACE2018}
\'{E}douard Bonnet and F.~Sikora.
\newblock The {PACE} 2018 parameterized algorithms and computational
  experiments challenge: The third iteration, 2018.
\newblock Available at \url{https://pacechallenge.org/files/PACE18-report.pdf}.

\bibitem{GabowS86}
H.~N. Gabow and M.~F.~M. Stallmann.
\newblock An augmenting path algorithm for linear matroid parity.
\newblock {\em Combinatorica}, 6(2):123--150, 1986.

\bibitem{guo:fvs}
J.~Guo, J.~Gramm, F.~H{\"u}ffner, R.~Niedermeier, and S.~Wernicke.
\newblock Compression-based fixed-parameter algorithms for feedback vertex set
  and edge bipartization.
\newblock {\em J. Comput. Syst. Sci.}, 72(8):1386--1396, 2006.

\bibitem{IwataGithub}
K.~Imanishi and Y.~Iwata.
\newblock Feedback {V}ertex {S}et solver, entry to {PACE} 2016, 2016.
\newblock \texttt{http://github.com/wata-orz/fvs}.

\bibitem{Iwata17}
Y.~Iwata.
\newblock Linear-time kernelization for feedback vertex set.
\newblock In I.~Chatzigiannakis, P.~Indyk, F.~Kuhn, and A.~Muscholl, editors,
  {\em 44th International Colloquium on Automata, Languages, and Programming,
  {ICALP} 2017, July 10-14, 2017, Warsaw, Poland}, volume~80 of {\em LIPIcs},
  pages 68:1--68:14. Schloss Dagstuhl - Leibniz-Zentrum fuer Informatik, 2017.

\bibitem{IwataWY16}
Y.~Iwata, M.~Wahlstr{\"{o}}m, and Y.~Yoshida.
\newblock Half-integrality, {LP}-branching, and {FPT} algorithms.
\newblock {\em {SIAM} J. Comput.}, 45(4):1377--1411, 2016.

\bibitem{IwataYY17}
Y.~Iwata, Y.~Yamaguchi, and Y.~Yoshida.
\newblock Linear-time {FPT} algorithms via half-integral non-returning a-path
  packing.
\newblock {\em CoRR}, abs/1704.02700, 2017.

\bibitem{fvs6}
I.~A. Kanj, M.~J. Pelsmajer, and M.~Schaefer.
\newblock Parameterized algorithms for feedback vertex set.
\newblock In R.~G. Downey, M.~R. Fellows, and F.~K. H.~A. Dehne, editors, {\em
  IWPEC}, volume 3162 of {\em Lecture Notes in Computer Science}, pages
  235--247. Springer, 2004.

\bibitem{karp}
R.~M. Karp.
\newblock Reducibility among combinatorial problems.
\newblock In R.~E. Miller and J.~W. Thatcher, editors, {\em Complexity of
  Computer Computations}, The IBM Research Symposia Series, pages 85--103.
  Plenum Press, New York, 1972.

\bibitem{our-repo}
K.~Kiljan and M.~Pilipczuk.
\newblock Experimental evaluation of parameterized algorithms for {F}eedback
  {V}ertex {S}et: code repository, 2018.
\newblock \texttt{https://bitbucket.org/marcin\_pilipczuk/fvs-experiments}.

\bibitem{KociumakaP}
T.~Kociumaka and M.~Pilipczuk.
\newblock Faster deterministic feedback vertex set.
\newblock {\em Inf. Process. Lett.}, 114(10):556--560, 2014.

\bibitem{PACE2016tests}
C.~Komusiewicz.
\newblock {PACE} 2016: tests for {F}eedback {V}ertex {S}et, 2016.
\newblock \texttt{https://github.com/ckomus/PACE-fvs}.

\bibitem{PilipczukPACE}
M.~Pilipczuk.
\newblock Feedback {V}ertex {S}et solver, entry to {PACE} 2016, 2016.
\newblock \texttt{http://bitbucket.com/marcin\_pilipczuk/fvs-pace-challenge}.

\bibitem{fvs5}
V.~Raman, S.~Saurabh, and C.~R. Subramanian.
\newblock Faster fixed parameter tractable algorithms for finding feedback
  vertex sets.
\newblock {\em ACM Transactions on Algorithms}, 2(3):403--415, 2006.

\bibitem{fvs:quadratic-kernel}
S.~Thomass{\'e}.
\newblock A $4k^2$ kernel for feedback vertex set.
\newblock {\em ACM Transactions on Algorithms}, 6(2):32:1--32:8, 2010.

\end{thebibliography}

\end{document}